\documentclass[prd,twocolumn,nofootinbib]{revtex4}
\usepackage{graphicx, epsfig}
\usepackage{color}
\usepackage{mathrsfs}
\usepackage {amssymb}
\usepackage {amsmath}

\newcommand{\nc}{\newcommand}

\nc{\ba}{\begin{eqnarray}}
\nc{\ea}{\end{eqnarray}}
\newcommand\be{\begin{equation}}
\newcommand\beq{\begin{equation}}
\newcommand\ee{\end{equation}}
\newcommand\eeq{\end{equation}}

\nc{\ga}{\gamma}
\nc{\tnu}{\bar{\nu}}
\nc{\tmu}{\bar{\mu}}
\nc{\tq}{\tilde{q}}

\nc{\x}{{\bf{x}}}
\nc{\bmi}{\bar \mu_{i} }
\nc{\bm}{\bar \mu }
\newcommand{\gsim}{\raise.3ex\hbox{$>$\kern-.75em\lower1ex\hbox{$\sim$}}}
\newcommand{\lsim}{\raise.3ex\hbox{$<$\kern-.75em\lower1ex\hbox{$\sim$}}}
\nc{\gm}{\gamma }
\nc{\PIJ}{P_{,X_{IJ}}}
\nc{\hP}{\hat{P}}
\nc{\PX}{P_{,X}}
\nc{\half}{\frac{1}{2}}
\nc{\p}{\phi}
\newcommand{\dn}[2]{{\mathrm{d}^{{#1}}{{#2}}}}
\nc{\s}{\sigma}

\newcommand\bk{\boldsymbol{k}}

\nc\bfphi{{\bf \phi}}

\def\tP{{\tilde P}}
\def\tX{{\tilde X}}
\def\tG{{\tilde G}}

\def\D{{\cal D}}
\def\R{{\cal R}}
\def\A{{\cal A}}

\input{epsf.sty}

\begin{document}

\title{Primordial fluctuations and non-Gaussianities in multi-field DBI inflation}

\author{David Langlois$^1$, S\'ebastien Renaux-Petel$^1$, Dani\`ele A.~Steer$^{1,2}$ and Takahiro Tanaka$^3$}

\affiliation{$^1$APC, UMR 7164, 10 rue Alice Domon et L\'eonie Duquet,75205 Paris Cedex 13, France}
\affiliation{$^2$CERN Physics Department, Theory Division, CH-1211 Geneva 23, Switzerland}
\affiliation{$^3$Yukawa Institute for Theoretical Physics, Kyoto University, Kyoto, 606-8502, Japan.}

\date{\today}

\begin{abstract}
We study Dirac-Born-Infeld (DBI) inflation models with multiple scalar fields. We show that the adiabatic and entropy modes propagate with a common effective sound speed  and are thus amplified at the sound horizon crossing. In the small sound speed limit, we find that the amplitude of the entropy modes is much higher than that of the adiabatic modes. We show that this could strongly affect the observable curvature power spectrum as well as the amplitude of non-Gaussianities, although their shape remains as in the single-field DBI case. 
\end{abstract}

\maketitle

The last decade has seen an accumulation of cosmological data of increasing precision.  Together with future experiments planned to measure the CMB fluctuations with yet further accuracy,  we may be able to piece together more clues about early universe physics.
In parallel with this observational effort, there has been tremendous progress in recent years in the construction of early universe models in the framework of high energy physics and string theory.

A particularly interesting class of models based on string theory is known as DBI inflation \cite{st03,ast04}, associated with the motion of a D3-brane in a higher-dimensional background spacetime. 
The characteristic of DBI inflation, and that which gives it its name, is that the action is of the Dirac-Born-Infeld (DBI) type and thus contains non-trivial kinetic terms. Most studies of DBI inflation models (or even of string based inflationary models) have so far concentrated on a {\it single-field} description meaning, in the DBI case, that the inflaton corresponds to a radial coordinate of the brane in the extra dimensions.  Taking into account the ``angular'' coordinates of the brane naturally leads to a {\it multi-field} description since each brane coordinate in the extra dimensions gives rise to a scalar field from the effective four-dimensional point of view.  This setup has started to be explored only very recently \cite{Easson:2007dh,Huang:2007hh}.

In this Letter, we show that the {\it multi-field} DBI action contains some terms, higher order in space-time gradients and vanishing in the homogeneous case, 
which have been overlooked.  The inclusion of these terms leads to drastic consequences on the primordial fluctuations generated in these types of models. The scalar-type perturbations in multi-field models can be divided into (instantaneous) adiabatic modes, fluctuations along the trajectory in field space, and entropy modes which are orthogonal to the former \cite{Gordon:2000hv}. In contrast with previous expectations, we show that in DBI models, these two classes of modes propagate with the {\it same} speed, namely an effective speed of sound $c_s$ smaller than the speed of light. As a consequence, the amplification of quantum fluctuations occurs at the sound horizon crossing for both types of modes. 
Moreover, when $c_s\ll 1$, this leads to an {\it enhancement} of the amplitude of the entropy modes with respect to that of the usual adiabatic modes.  As primordial non-Gaussianities --- potentially detectable in forthcoming experiments if strong enough --- discriminate between various models, we also study the impact of the entropy modes on non-Gaussianity in the DBI case.

Our starting point is the DBI Lagrangian governing the dynamics of a D3-brane,
\beq
L_{DBI}=-\frac{1}{f}\sqrt{-\det\left(g_{\mu\nu}+f G_{IJ}\partial_\mu\phi^I\partial_\nu \phi^J\right)} \,,
\label{DBI}
\eeq
where $f=f(\phi^I)$ is a function of the scalar fields $\phi^I$ ($I=1,2,\ldots$), and $G_{IJ}(\phi^K)$ is a metric in field space.   From a  
higher-dimensional point of view, (\ref{DBI}) is proportional to the square root of the determinant of the induced metric on the brane, meaning that the $\phi^I$ correspond to the brane coordinates in the extra dimensions, $f$ embodies the warp factor, and 
$G_{IJ}$ is (up to a scale factor) the metric in the extra dimensions.  We also allow for the presence of a potential and hence consider a full action of the form 
\ba
\label{action1}
S &=&  \int {\rm d}^4 x \sqrt{-g}\left[\frac{{\cal R}}{2}   +  P\right] , 
\nonumber
\\
P&=&-\frac{1}{f(\bfphi^I)}\left(\sqrt{{\cal D}}-1\right) -V(\bfphi^I)\,,
\ea
where we have set $8\pi G=1$.  The determinant 
${\cal D}=  \det(\delta^{\mu}_{\nu}+f G_{IJ}\partial^{\mu} \p^I \partial_{\nu} \p^J )$ coming from Eq.~(\ref{DBI}) can be rewritten as
\ba
\label{D}
{\cal D}&=&\det(\delta_{I}^{\, J}-2f X_{I}^{\, J})
\nonumber
\\
& =& 1-2f G_{IJ}X^{IJ} + 4f^2 X^{[I}_IX_J^{J]} 
\nonumber
\\ &&-8f^3 X^{[I}_IX_J^{J} X_K^{K]}+16f^4 X^{[I}_IX_J^{J} X_K^{K}X_L^{L]},
\ea
where we have defined 
\beq
X^{IJ}\equiv -\frac{1}{2} \partial^{\mu} \p^I \partial_{\mu} \p^J , \quad X_I^J=G_{IK}X^{KJ},
\eeq
and where the brackets denote antisymmetrisation of the field indices.  In the {\it single-field} case, $I=1$, 
the terms in $f^2$, $f^3$ and $f^4$ in (\ref{D}) vanish.
This is also true for {\it multiple homogeneous} scalar fields for which ${X}^{IJ}=\frac{1}{2}\dot\phi^I\dot\phi^J$. However, for {\it multiple inhomogeneous} scalar fields, these terms, which are higher order in gradients and have not been considered in previous works, 
do not vanish.  We now show that they change drastically the behaviour of perturbations.

In order to study the dynamics of linear perturbations about a homogeneous cosmological solution, we expand the initial action (\ref{action1}) to second order in the linear perturbations, including both metric and scalar field perturbations. This is a constrained system, and the number of (scalar) degrees of freedom is the same as the number of scalar fields. It is convenient to express these degrees of freedom in terms of the scalar perturbations defined in the flat gauge, usually denoted $Q^I$. 
To obtain the second-order action, we follow the procedure outlined in \cite{ds} for a 
Lagrangian of the form $P(X,\phi^J)$, with $X=G_{IJ}X^{IJ}$:  as we have stressed above the multi-field DBI Lagrangian is {\it not} of this form, but despite that it can be rewritten as
\be
\tilde{P}(\tilde X,\phi^K) = -\frac{1}{f} \left(\sqrt{1-2f\tilde{X}}-1 \right) - V\,,
\label{Ptildedef}
\ee
where $\tilde{X}=(1-\D)/(2f)$.
Although in the homogeneous background 
$\tilde{X}$ and $X$ coincide, their perturbed values differ.  Taking into account the corresponding extra terms, one can show \cite{preparation} that the second-order action can be written in the compact form
\begin{eqnarray}
S_{(2)}&=&\half \int {\rm d}t \,\dn{3}{x}\,    a^3\left[ \tP_{,\tX}\tG_{IJ}
 \mathcal{D}_t Q^I \mathcal{D}_t Q^J  \right.
 \cr
 &&
\qquad -\frac{c_s^2}{a^2}\tP_{,\tX}\tG_{IJ} \partial_i Q^I \partial^i Q^J 
 \cr
  && 
  \left.
  - {\cal M}_{IJ}Q^I Q^J + 2 \tP_{,\tX J} \dot \p_I Q^J \mathcal{D}_t Q^I  \right]\,.
\label{2d-order-action}
\end{eqnarray}
Here $a$ is the scale factor; the effective (squared) 
mass matrix is 
\begin{eqnarray}
 {\cal M}_{IJ}  &=& -\mathcal{D}_I \mathcal{D}_J \tP - \tP_{,\tX} \mathcal{R}_{IKLJ}\dot \p^K \dot
\p^L
\nonumber\\
&+&\frac{X \tP_{,\tX}}{H} (\tP_{,\tX J}\dot \p_I+\tP_{,\tX I}\dot \p_J) 
 + \frac{\tX \tP_{,\tX}^3}{2 H^2}(1-\frac{1}{c_s^2})\dot \p_I \dot \p_J
 \nonumber\\
&-& \frac{1}{a^3}\mathcal{D}_t\left[\frac{a^3}{2H}\tP_{,\tX}^2\left(1+\frac{1}{c_s^2}\right)\dot \p_
I \dot \p_J\right]  \,,
\label{Interaction matrix}
\end{eqnarray}
and we have introduced covariant derivatives $ \mathcal{D}_I$ defined with respect to the field space metric $G_{IJ}$, as well as the time covariant derivative  $\mathcal{D}_t Q^{I} = \dot{Q}^I + \Gamma^{I}_{JK} \dot{\phi}^J Q^{K}$  where $\Gamma^{I}_{JK} $ is the Christoffel symbol constructed from $G_{IJ}$ and $\mathcal{R}_{IKLJ}$ is the corresponding Riemann tensor.  Finally, we have defined  the (background) matrix
\beq
\tilde{G}_{IJ}=G_{IJ}+\frac{2fX}{1-2fX} e_{\s I} e_{\s J}=\perp_{IJ}+\frac{1}{c_s^2} e_{\s I} e_{\s J}\,,
\eeq
where $e_{\s}^I=\dot\phi^I/\sqrt{2X}$ ($\dot \s \equiv \sqrt{2X}$ is also used in the following) is the unit vector pointing along the trajectory in field space, $\perp_{IJ}\equiv
G_{IJ}-e_{\s I} e_{\s J}$ is  the projector orthogonal to the vector $e_{\s}^I$,
and 
\beq
c_s^2 \equiv \frac{\tP_{,\tX}}{\tP_{,\tX}+2\tX \tP_{,\tX\tX}}=1-f \dot{\sigma}^2.
\eeq

Let us stress that the only difference between action (\ref{2d-order-action}) and the corresponding expression in \cite{ds} is the term in spatial gradients, with coefficient $ c_s^2 \tilde{P}_{,\tX} \tilde{G}_{IJ}$ instead of $\tilde{P}_{,X} G_{IJ}$.  This crucial difference shows that  {\it all} perturbations, both adiabatic and entropic, propagate with the {\it same} speed of sound in multi-field DBI inflation, in contrast with \cite{ds,Easson:2007dh,Huang:2007hh} where they have different speeds.
Finally, one should recall that the above expressions apply to the DBI context where
$\tilde{P}$ is given in (\ref{Ptildedef}) so that $\tP_{,\tX} = 1/c_s.$

For simplicity, let us now restrict our attention to two fields ($I=1,2$). The perturbations can then be decomposed into $Q^I=Q_\s e^I_\s+Q_s e^I_s$\,, where $e_s^I$, the unit vector orthogonal to $e^I_\s$, characterizes the entropy direction. (For $N$ fields, the entropy modes would span an $(N-1)$-dimensional subspace in field space.) 
As in standard inflation, it is more convenient, after going to conformal time $\tau = \int {{\rm d}t}/{a(t)}$, to work in terms of the canonically normalized fields
\be
v_{\s}\equiv \frac{a}{c_s} \sqrt{\tP_{,\tX}}\, Q_{\s} \,,\qquad \,v_{s}\equiv a\,\sqrt{\tP_{,\tX}}\, Q_s\,.
\label{v}
\ee
Note that the adiabatic and entropy coefficients differ because $\tilde{G}_{IJ}$ is anisotropic. The equations of motion for $v_{\s}$ and $v_s$ then follow from the action (\ref{2d-order-action}), reexpressed in terms of the rescaled quantities (\ref{v}). One finds 
\begin{eqnarray}
v_{\s}''-\xi v_{s}'+\left(c_s^2 k^2-\frac{z''}{z}\right) v_{\s} -\frac{(z \xi)'}{z}v_{s}&=&0\,, \qquad
\label{eq_v_sigma}
\\
v_{s}''+\xi  v_{\s}'+\left(c_s^2 k^2- \frac{\alpha''}{\alpha}+a^2\mu_s^2\right) v_{s} - \frac{z'}{z} \xi v_{\s}&=&0\,,
\label{eq_v_s}
\end{eqnarray}
where 
\begin{eqnarray}
\label{11}
&&\xi=\frac{a}{\dot \s \tP_{,\tX} c_s}[(1+c_s^2)\tP_{,s}-c_s^2 \dot \s^2 \tP_{,\tX s}]\,,
\\
&&z=\frac{a \dot \s }{c_s H}\sqrt{\tP_{,\tX}}, \qquad \alpha=a\sqrt{\tP_{,\tX}}\,,
\label{z,a}
\end{eqnarray}
and $\mu_s^2$ follows from the mass matrix (\ref{Interaction matrix}) (see \cite{ds,preparation} for details).
We will assume that the effect of the coupling $\xi$ can be neglected when the scales of interest cross out the {\it sound horizon}, so that the two degrees of freedom are decoupled and the system can easily be quantized. In the {\it slow-varying} regime, where the time evolution of $H$, $c_s$ and $\dot \s$ is small with respect to that of the scale factor, one gets $z''/z\simeq 2/\tau^2$ and $\alpha''/\alpha\simeq 2/\tau^2$. 
The  solutions of (\ref{eq_v_sigma}) and (\ref{eq_v_s}) corresponding to the Minkowski-like  vacuum on small scales are thus
\beq
v_{\s\, k}\simeq v_{s\, k}\simeq  \frac{1}{\sqrt{2k c_s}}e^{-ik c_s \tau }\left(1-{i\over k c_s\tau}\right),
\eeq
when $\mu^2_s/H^2$ is negligible for the entropic modes (if $\mu^2_s/H^2$ is large the entropic modes are suppressed).
The power spectra for $v_\s$ and $v_s$ after sound horizon crossing therefore have the same amplitude ${\cal P}_{v} =(k^3/2\pi^2)|v_{k}|^2$. The power spectra for $Q_\s$ and $Q_s$ are thus
\beq
\label{power_sigma}
{\cal P}_{Q_\s}\simeq\frac{H^2}{4\pi^2 c_s \tP_{,\tX}}, \quad {\cal P}_{Q_s}\simeq\frac{H^2}{4\pi^2 c_s^3 \tP_{,\tX}}\,,
\eeq
evaluated at sound horizon crossing.
One recognizes the familiar result of k-inflation for the adiabatic part \cite{ArmendarizPicon:1999rj,Garriga:1999vw}, while for small $c_s$, the entropic modes are {\it amplified} with respect to the adiabatic modes:
$Q_s\simeq Q_{\sigma}/c_s$.

These results can be reexpressed in terms of the comoving curvature perturbation 
${\cal R}=(H/\dot \s)Q_{\s}\,$ with which it is useful to relate the perturbations during inflation to the primordial fluctuations during the standard radiation and present era. 
We recover the usual {\it single-field} result for the power spectrum of 
$\R$ at sound horizon crossing:
\be
{\cal P}_{\cal R_*}\simeq\frac{H^4}{8\pi^2 c_s \tX \tP_{,\tX}}=\frac{H^2}{8\pi^2 \epsilon c_s }\,,
\label{power-spectrum-R}
\ee
where $\epsilon=-\dot H / H^2 = \tP_{,\tX}\tX/H^2$ (the subscript $*$ indicates that the corresponding quantity is evaluated at sound horizon crossing). It is then convenient to define an entropy perturbation ${\cal S}=c_s\frac{H}{\dot \s}Q_{s}$ such that $ {\cal P}_{\cal S_*} \simeq {\cal P}_{\cal R_*}$.
The power spectrum for the tensor modes is, as usual, governed by the transition at {\it Hubble radius} and its amplitude, ${\cal P}_{\cal T}=(2H^2/\pi^2)_{k=aH}$,
is much smaller than the curvature amplitude for $c_s \ll 1$.

Leaving aside the possibility that the entropy modes during inflation lead directly to primordial entropy fluctuations that could be detectable in the CMB fluctuations (potentially correlated with adiabatic modes as discussed 
in \cite{Langlois:1999dw}), we consider here only the influence of the entropy modes on the final curvature perturbation. 
Indeed, on large scales, the curvature perturbation can evolve in time in the multi-field case, because of the entropy modes. This transfer from the entropic to the adiabatic modes depends on the details of the scenario and of the background trajectory in field space, but it can be parametrized by a transfer coefficient \cite{Wands:2002bn}
which appears in the formal solution $\R=\R_*+T_{ {\cal R}  {\cal S} } \cal S_*$ of the first-order evolution equations for $\cal R$ and $\cal S$ which follow from (\ref{eq_v_sigma}), (\ref{eq_v_s}) in the slow-varying regime on large scales.

This implies in particular that the final curvature power spectrum can be formally expressed as 
${\cal P}_{\cal R}=(1+T_{{\cal R} {\cal S}}^2) {\cal P}_{\cal R_{*}}$.  Let us define the ``transfer angle'' $\Theta$ ($\Theta=0$ if there is no transfer and $|\Theta|=\pi/2$ if the final curvature perturbation is mostly of entropic origin) by
\be
{\sin} \Theta =\frac{T_{ {\cal R}  {\cal S} }}{\sqrt{1+T^2_{ {\cal R}  {\cal S} }}}\,,
\label{correlation-result}
\ee
so that the curvature power spectrum at sound horizon crossing and  its observed value are related by
${\cal P}_{\cal R_{*}}={\cal P}_{\cal R} {\rm cos^2} \Theta$.
Finally the tensor to scalar ratio is given by
\be
 r \equiv \frac{{\cal P}_{\cal T}}{{\cal P}_{\cal R}}=16 \epsilon c_s {\rm cos^2} \Theta\,.
\label{tensor-to-scalar}
\ee
This expression combines the result of k-inflation, where the ratio is suppressed by the sound speed $c_s$ and of standard multi-field inflation \cite{Gordon:2000hv}.

We finally turn to primordial non-Gaussianities, whose detection would provide an additional window on the very early universe. This aspect is especially important for DBI models since it is well known that 
(single-field) DBI inflation  produces a (relatively) high level of non-Gaussianity for small $c_s$ \cite{ast04}. How, therefore, do the entropic modes, whose amplitude is much larger than that of the adiabatic fluctuations, affect the primordial non-Gaussianity? In the small $c_s$ limit, one can estimate the dominant contribution by extracting from the third-order Lagrangian the analogue of the terms giving the dominant contribution in the single-field case, but including now the entropy components.  These terms are \cite{preparation}
\ba
&& S_{(3)}
^{(\rm main)}
=\int {\rm d}t {\rm d}^3x \frac{a^3}{2 c_s^5 \dot \s}\left[\dot Q_{\s}^3+c_s^2 \dot Q_{\s}  \dot Q_{s}^2\right]
\cr
&&
 - \frac{a}{2 c_s^3 \dot \s}\left[ \dot Q_{\s} (\nabla  Q_{\s} )^2 -c_s^2 \dot{Q_{\s} }(\nabla  Q_{s} )^2+2 c_s^2 \dot {Q_s}\nabla Q_{\s} \nabla Q_s\right],
 \nonumber
\ea
where we have replaced $f$ by $1/\dot\sigma^2$ since, for $c_s\ll 1$, $f\dot\sigma^2\simeq 1$. 
Following the standard procedure \cite{Maldacena:2002vr,Seery:2005wm,Seery:2005gb} one can compute the 3-point functions involving adiabatic and entropy fields.  
The purely adiabatic 3-point function is naturally the same as in single-field DBI \cite{Chen:2005fe,Chen:2006nt}.  The new contribution is\begin{eqnarray}
&&\langle Q_{\s} (\boldsymbol{k}_1) Q_{s} (\boldsymbol{k}_2) Q_s (\boldsymbol{k}_3)\rangle  
\cr
&=&- (2 \pi)^3 \delta (\sum_i \bk_i) \frac{H^4}{4\sqrt{2 c_s \epsilon} c_s^4}\frac{1}{\left(\prod_i k_i^3\right) K^3} 
\left[ -2 k_1^2 k_2^2 k_3^2
\right.
\cr
&&\left. 
- k_1^2 (\bk_2 \cdot \bk_3)(2 k_2 k_3 -k_1 K +2 K^2)
\right.
\cr
&&\left. 
+k_3^2 (\bk_1 \cdot \bk_2)(2 k_1 k_2 -k_3 K +2 K^2) 
\right.
\cr
&&\left.
+k_2^2 (\bk_1 \cdot \bk_3)(2 k_1 k_3 -k_2 K +2 K^2) \right]\,,
\label{new-shape}
\end{eqnarray}
where $K=\sum_i k_i$.

We now relate the correlation functions of the scalar fields to the 3-point function of the curvature perturbation $\R$ which is the observable quantity.  It follows directly from above that
\begin{eqnarray}
&&{\cal R} \approx \A_\sigma  Q_{\s*} + \A_s  Q_{s*} 
\label{transfer}
\cr
&& \A_\sigma = \left( \frac{H}{\dot \s}\right)_*  \quad
\A_s =  T_{{\cal R} {\cal S}} \left( \frac{ c_s H}{\dot \s}\right)_*\,.
\label{comparision-transfer}
\end{eqnarray}
Hence, for vectors $\bk_i$ whose norms have the same order of magnitude (so that the slowly varying background parameters are evaluated at about the same time)
\begin{eqnarray}
&&\langle\R (\boldsymbol{k}_1)  \R (\boldsymbol{k}_2)  \R (\boldsymbol{k}_3)\rangle = (\A_{\s})^3\langle Q_{\s}  (\boldsymbol{k}_1) Q_{\s}  (\boldsymbol{k}_2)  Q_{\s}  (\boldsymbol{k}_3) \rangle
 \cr
  && 
+\A_{\s}(\A_{s})^2[\langle Q_{\s}  (\boldsymbol{k}_1)  Q_s (\boldsymbol{k}_2)  Q_s (\boldsymbol{k}_3)\rangle+  {\rm perm.}]
\cr 
&&= 
(\A_{\s})^3\langle Q_{\s}  (\boldsymbol{k}_1) Q_{\s}  (\boldsymbol{k}_2)  Q_{\s} (\boldsymbol{k}_3) \rangle
\left(1+T_{{\cal R} {\cal S}}^2\right).
\label{zeta-3}
\end{eqnarray}
As we see, the above quantity depends on the symmetrized version of the 3-point function (\ref{new-shape}), which has 
exactly the same shape as in single-field DBI.  
Note that the enhancement of the mixed correlation $\langle Q_{\s}  Q_{s} Q_s \rangle$ by a factor of $1/c_s^2$ is compensated by the ratio between $\A_{\s}$ and $\A_{s}$ so that the adiabatic and mixed  contributions in (\ref{zeta-3}) are exactly of the same order.
In principle, there are other contributions to the observable 3-point function, in particular those coming from the 4-point function of the scalar fields, which can be reexpressed in terms of the power spectrum via Wick's theorem \cite{Lyth:2005fi}. The amplitude of this contribution will depend on the specific models.  We implicitly ignore them in the following.

The non-Gaussianity parameter $f_{NL}$  is defined by
\ba
\langle\R (\boldsymbol{k}_1)  \R (\boldsymbol{k}_2)  \R (\boldsymbol{k}_3)\rangle = \qquad \qquad \qquad
\nonumber
\\
-(2 \pi)^7 \delta (\sum_i \bk_i) \left[ \frac{3}{10}f_{NL}({\cal P}_{\cal R})^2\right] \frac{\sum_i k_i^3}{\prod_i k_i^3}\,,
\label{def f_NL}
\ea
from which we obtain, for the equilateral configuration,
\be
f_{NL}^{(3)}=-\frac{35}{108}\frac{1}{c_s^2}\frac{1}{1+T^2_{{\cal R} {\cal S}} }=-\frac{35}{108}\frac{1}{c_s^2} {\cos^2} \Theta \,.
\label{f_NL3}
\ee
One can easily understand this result.  The curvature power spectrum is amplified by a factor of $(1+T^2_{{\cal R} {\cal S}})$ due to the feeding of curvature by entropy modes.
Similarly the 3-point correlation function for ${\cal R}$ resulting from the 3-point correlation functions of the adiabatic and entropy modes is enhanced by the same factor $(1+T^2_{{\cal R} {\cal S}})$. However, since $f_{NL}$ is roughly the ratio of the 3-point function with respect to the {\it square} of the power spectrum, one sees that $f_{NL}$ is now {\it reduced} by the factor 
$(1+T^2_{{\cal R} {\cal S}})$. The so-called UV model of DBI inflation is under strong observational pressure because it generates a high level of non-Gaussianities that exceed the experimental bound \cite{Bean:2007hc,Peiris:2007gz}. We stress that their reduction by multiple-field effects may be very important for
model-building. 

We end by revisiting the consistency condition relating  the non-Gaussianity of the curvature perturbation, the tensor to scalar ratio $r$, and the tensor spectral index $n_{\cal T} =-2\epsilon$, given in \cite{Lidsey:2006ia} for single-field DBI.  In our case, substituting $f_{NL}^{(3)} \simeq -\frac{1}{3}\frac{1}{c_s^2} {\cos^2} \Theta$ in (\ref{tensor-to-scalar}), gives
\be
r+8 n_{\cal T}=-r\left( \sqrt{-3 f_{NL}^{(3)}}\cos^{-3} \Theta-1\right),
\label{consistency}
\ee
As we can can see from (\ref{f_NL3}) and (\ref{consistency}), violation of the standard inflation consistency relation (corresponding to a vanishing right-hand side in (\ref{consistency})) would be stronger in multi-field DBI than in single-field DBI, and thus easier to detect.  In the multi-field case the consistency condition is only an inequality (unless $\Theta$ is observable when the entropy modes survive after inflation) from which we can infer the transfer angle.  

To summarize, we have shown that  both adiabatic and entropy modes propagate with the same speed of sound $c_s$, in multi-field DBI models. Both modes are thus amplified at the sound horizon crossing, with an enhancement of the entropy modes with respect to the adiabatic ones in the small $c_s$ limit. The amplitude of the non-Gaussianities, which are important in DBI models, is also strongly affected by the entropy modes, although their shape remains as in the single-field case. 
All these features are generic  in any model governed by the multi-field DBI action.  The model-specific quantity (depending on the field metric, the warp factor and the potential) is the transfer coefficient between the initial entropy modes and the final curvature perturbation between the time when the fluctuations cross out the sound horizon and the end of inflation. Recent analyses
\cite{Lalak:2007vi,Brandenberger:2007ca} in slightly different contexts show that this transfer can be very efficient, leading to a final curvature perturbation of entropic origin (as in the curvaton scenario).
More generally, our results show that multi-field effects, common in string theory motivated inflation models, deserve close attention as the entropy modes produced could significantly affect the cosmological observable quantities.

{\it Acknowledgement}. We are grateful for a CNRS-JSPS grant.


\end{document}